\documentclass{emulateapj}
\usepackage{amsmath}
\usepackage{ulem}
\usepackage{CJK}

\newcommand\bld[1]{\mbox{\boldmath $#1$}}
\newcommand{\del}{\partial}

\shorttitle{GRB/Blazar polarization} \shortauthors{Deng et al.} \slugcomment{}
\begin{document}
\begin{CJK*}{UTF8}{gbsn}
\title{Collision-induced magnetic reconnection and a unified interpretation of polarization properties of GRBs and Blazars}
\author{Wei Deng (邓巍)\altaffilmark{1,2}, Haocheng Zhang (张昊成)\altaffilmark{3,2}, Bing Zhang (张冰)\altaffilmark{1}, Hui Li (李晖)\altaffilmark{2}}

\affil{\altaffilmark{1}Department of Physics and Astronomy, University of Nevada Las Vegas, Las Vegas, NV 89154, USA\\deng@physics.unlv.edu, zhang@physics.unlv.edu}
\affil{\altaffilmark{2}Los Alamos National Laboratory, Los Alamos, NM 87545, USA\\ hli@lanl.gov}
\affil{\altaffilmark{3}Astrophysical Institute, Department of Physics and Astronomy, Ohio University, Athens, OH 45701, USA\\hz193909@ohio.edu}

\begin{abstract}
The jet composition and energy dissipation mechanism of Gamma-ray bursts (GRBs) and Blazars are fundamental questions which remain not fully understood. One plausible model is to interpret the $\gamma$-ray emission of GRBs and optical emission of blazars as synchrotron radiation of electrons accelerated from the collision-induced magnetic dissipation regions in Poynting-flux-dominated jets. The polarization observation is an important and independent information to test this model. Based on our recent 3D relativistic MHD simulations of collision-induced magnetic dissipation of magnetically dominated blobs, here we perform calculations of the polarization properties of the emission in the dissipation region and apply the results to model the polarization observational data of GRB prompt emission and blazar optical emission. We show that the same numerical model with different input parameters can reproduce well the observational data of both GRBs and blazars, especially the $90^{\circ}$ polarization angle (PA) change in GRB 100826A and the $180^{\circ}$ PA swing in Blazar 3C279. This supports a unified model for GRB and blazar jets, suggesting that collision-induced magnetic reconnection is a common physical mechanism to power the relativistic jet emission from events with very different black hole masses.
\end{abstract}


\section{Introduction\label{sec:intro}}

Gamma-ray bursts (GRBs) are the most luminous explosions in the universe, which mark ultra-relativistic jets launched from new-born stellar mass black holes that beam towards earth \citep{kumarzhang15}. Blazars are moderately relativistic jets launched from super-massive black holes towards earth \citep{ghisellini15}. 
The energy composition of these relativistic jets, which is still in debate, is a fundamental property that defines the subsequent energy dissipation process, particle acceleration mechanism, and radiation mechanism, which are directly connected to the observational properties such as light curves, spectra, and polarization characteristics. An important parameter to classify relativistic jets is the magnetization parameter,  $\sigma= {E_{\rm em}}/{h}$, where $h=\rho c^2+\hat{\gamma}P/(\hat{\gamma}-1)$ is the specific enthalpy defined in the fluid rest frame, $\rho$ is the rest mass density, $P$ is the gas pressure, $\hat{\gamma}$ is the adiabatic index, and $E_{\rm em}$ is the EMF energy density calculated by $E_{\rm em}=(\bld{B_0}^2+\bld{E_0}^2)/8 \pi$ in the fluid rest frame of each simulation cell. A jet is matter-flux-dominated (MFD) when $\sigma<1$ \citep{meszarosrees00,rees94,kobayashi97,daigne98}, and is Poynting-flux-dominated (PFD) when $\sigma > 1$ \citep{zhangyan11,lovelace76,blandford76,colgate04,giannios09}.

Recently, evidence suggesting that at least some GRBs may be produced through magnetic dissipation within moderately PFD jets has been accumulated, including the weak or the lack of bright thermal emission in the spectra of most GRBs \citep{zhangpeer09}, linearly polarized emission with high polarization degree (PD) in both GRB prompt emission \citep{yonetoku11} and early optical afterglow \citep{mundell13}, as well as the lack of associated high-energy neutrinos with GRBs \citep{zhangkumar13,aartsen15}. In the blazar field, polarized optical emission has been commonly observed \citep{abdo10,blinov15}. A connection between blazar jets and GRB jets has been established through several observational correlations \citep{nemmen12,zhangjin13,wu16}. Theoretically, a possible connection between jets in different scales has been proposed \citep{mirabel04,zhangsn07}, but no detailed studies have been carried out to show that the same physical model can account for the observational data in both GRBs and blazars.

Within the context of GRBs, \cite{zhangyan11} suggested that collision-induced magnetic reconnection and turbulence (ICMART) can be an efficient mechanism to dissipate energy in a moderately PFD outflow, and the model can overcome some difficulties in the traditional internal shock models and account for many GRB observations. Recently, we \citep{deng15} carried out the first detailed numerical study on collision-induced magnetic dissipation, and demonstrated that such processes can indeed dissipate magnetic energy with a high efficiency, and that local, relativistically-boosted regions due to high-$\sigma$ reconnections can indeed form, which mimic the so-called ``mini-jets'' invoked in interpreting the rapid variabilities in GRBs \citep{narayan09,zhangyan11} and blazars \citep{giannios09}. It is then encouraging to extend our numerical model to study radiation from the jets, and directly compare the model results with observations.

In this paper, we study radiation properties of jets based on our 3D relativistic MHD numerical simulations of collision-induced magnetic dissipation model \citep{deng15}, paying special attention to the polarization properties of the jets. The observational motivations include the detection of high-degree of linear polarization as well as a $90^{\circ}$ polarization angle (PA) change between two major flares in the prompt $\gamma$-ray emission phase of GRB 100826A \citep{yonetoku11}, and the detections of a $180^{\circ}$ PA swing during strong flaring emission in some blazar sources, especially 3C279 \citep{abdo10,blinov15}. Several models \citep{lundman14,marscher08,marscher14,zhang15} have been proposed to interpret these phenomena within the GRB context or blazar context, respectively, but these models introduce very different assumptions, so that there is no unified understanding to the phenomena. 

\section{3D relativistic MHD simulation}

\subsection{MHD code introduction}

The 3D special relativistic MHD (SRMHD) code, that we are using here, is named as ``LA-COMPASS" \citep{lili03} which was first developed at Los Alamos National Laboratory. It uses the higher-order Godunov-type finite-volume methods to solve the ideal MHD equations as following:
\begin{eqnarray}
&\frac{\partial(\Gamma \rho)}{\partial t}+\nabla \cdot (\Gamma \rho \bld{V}) = 0,\\
&\frac{\partial}{\partial t}(\frac{\Gamma^2 h}{c^2}\bld{V}+\frac{\bld{E}\times \bld{B}}{4 \pi c})+
\nabla \cdot[\frac{\Gamma^2 h}{c^2}\bld{V}\otimes \bld{V} \nonumber \\
&+(p+\frac{B^2+E^2}{8 \pi})\bld{I}-\frac{\bld{E}\otimes \bld{E}+\bld{B}\otimes \bld{B}}{4 \pi}] = 0, \\
&\frac{\partial}{\partial t}(\Gamma^2 h-p-\Gamma \rho c^2 + \frac{B^2+E^2}{8 \pi}) \nonumber \\
&+ \nabla \cdot[(\Gamma^2 h-\Gamma \rho c^2)\bld{V} +\frac{c}{4 \pi}\bld{E}\times \bld{B}]=0,\\
&\frac{\partial \bld{B}}{\partial t}+ c \nabla \times \bld{E}=0,\\
&\bld{E}=-\frac{\bld{V}}{c}\times \bld{B},\\
&p=(\hat{\gamma}-1)u,
\end{eqnarray}
where $\bld{B}$, $\bld{E}$, and $\bld{V}$ are the vectors of magnetic field, electric field, and fluid velocity, respectively, $\Gamma$ and $u$ are the Lorentz factor and internal energy density, respectively (see more in \cite{deng15}).

\subsection{Problem set up}

We set up the two-blob collision problem similar to \cite{deng15}. We consider that the black hole central engine of a GRB or a blazar launches a PFD jet with both toroidal and poloidal magnetic field components. The jets are episodic, probably due to variations of the accretion rate, intrinsic episodic magnetic activities from the accretion disk \citep{yuanzhang12}, or current-driven kink instabilities during jet propagation \citep{guan14,mizuno14}, forming discrete high-$\sigma$ blobs with both toroidal and poloidal field components \citep{li06}. The collisions between the two blobs lead to discharge of magnetic energy through reconnection \citep{zhangyan11,deng15}, which lead to efficient particle acceleration \citep{guo14} and synchrotron radiation. We simulate the collisions in the jet co-moving frame, which is the center of mass frame of the two blobs.
The magnetic field configuration of each magnetic blob is initialized using the model from \cite{li06}. We introduce the equations in the cylindrical coordinates $(r, \phi, z)$ first, and then transfer them to the Cartesian coordinates in our simulations. From the center ($r=0$) of each blob, the field is assumed to be axisymmetric with the $r-$ and $z-$ components of the poloidal field:
\begin{equation}
B_{r} = -\frac{1}{r}\frac{\del \Phi}{\del z} = 2B_0\frac{zr}{r^2_{0}} \exp \left(-\frac{r^2+z^2}{r^2_{0}}\right),
\label{equ:Br}
\end{equation}
and
\begin{equation}
B_{z} = \frac{1}{r}\frac{\del \Phi}{\del r} = 2B_0 \left(1-\frac{r^2}{r^2_{0}}\right)\exp\left(-\frac{r^2+z^2}{r^2_{0}}\right),\label{equ:Bz}
\end{equation}
where, $B_0$ and $r_0$ are the normalization factor for the magnetic strength and the characteristic radius of the magnetic blob, respectively.
The poloidal field is closed, which keeps the net global poloidal flux zero. The toroidal field configuration is motivated by the consideration of a rapidly rotating central engine, whose spin shears the poloidal flux to form the toroidal flux. 
It has the form
\begin{equation}
B_{\phi}  = \frac{\alpha\Phi}{r_0 r} = B_0~ \alpha \frac{r}{r_0}\exp\left(-\frac{r^2+z^2}{r^2_{0}}\right)~,
\label{equ:B_phi}
\end{equation}
where the parameter $\alpha$ controls the toroidal-to-poloidal flux ratio. In our simulation, we set $\alpha=3$, which suggests that the two flux components are roughly equal to each other \citep{li06}.

The initial conditions for the GRB and blazar simulations are presented in Table \ref{tab:para}. The simulation box size is 10$L_0$ $\times$ 10$L_0$ $\times$ 10$L_0$ for the GRB case, and 10$L_0$ $\times$ 10$L_0$ $\times$ 20$L_0$ for the blazar case. The total simulation time is 160$t_0$ for both cases. We also assume a co-moving observer in the x-axis direction, and a bulk motion along the $z$-axis.
The bulk Lorentz factor is set to 300 for the GRB case and 20 for the blazar case, respectively. From the MHD simulations, we can roughly estimate the observed luminosity assuming that the radiation is isotropic in the jet comoving frame. For the GRB case, the average observed luminosity from 0 to 100 seconds is $\overline{L}_\Omega=\Gamma^4\cdot dP'/d\Omega'\approx \Gamma^4\cdot 5\times 10^{40}$ergs/s$~ \approx 4\times 10^{50}$ergs/s; For the blazar case, the average observed luminosity from 0 to 20 days is $\overline{L}_\Omega=\Gamma^4\cdot dP'/d\Omega'\approx \Gamma^4\cdot 6.5\times 10^{40}$ergs/s$~ \approx 10^{46}$ergs/s. Both values are typical for the respective systems.

\begin{table}
\scriptsize 
\caption{The parameters of our MHD simulations in the simulation (jet comoving) frame}
\begin{tabular}{|c||c|c|c|c|}
\hline
\multicolumn{1}{|c||}{Case} & \multicolumn{2}{c|}{GRB 100826A} & \multicolumn{2}{c|}{Blazar 3C279}\\\hline
Parameter & Code & Physical & Code & Physical\\\hline
$\sigma_c$ & 6 & 6 & 2 & 2\\
$\overline{B}$ & 0.25 & 750 (G) & 0.35 & 0.35 (G)\\
$\left|V_z\right|$ & 0.3 & 0.3c & 0.8 & 0.8c\\
$t_0$ & 1 & 375 (s) & 1 & $9.5\times 10^{5}$ (s)\\
$L_0$ & 1 & $1.13\times 10^{13}$ (cm) & 1 & $2.86\times 10^{16}$ (cm)\\
$x_s$ & 0.01 & $1.13\times 10^{11}$ (cm) & 1 & $2.86\times 10^{16}$ (cm)\\
$r_0$ & 2 & $2.25\times 10^{13}$ (cm) & 2.5 & $7.15\times 10^{16}$ (cm)\\
$\alpha$ & 3 & 3 & 3 & 3\\
$P$ & $10^{-2}$ & $9\times 10^{4}$ (ergs/cm$^3$) & $10^{-2}$ & $10^{-2}$ (ergs/cm$^3$)\\
$\rho_{\rm bkg}$ & $10^{-1}$ & $10^{-15}$ (g/cm$^3$) & $10^{-1}$ & $1.1\times 10^{-22}$ (g/cm$^3$)\\
\hline
\end{tabular}
\label{tab:para}
\renewcommand{\thefootnote}
\thefootnote{{\bf Notes.} For each parameter, we give the normalization relationship between the dimensionless code units and the physical units. Fig. 1 shows the model geometry. Here $\sigma_c$ is the blob $\sigma$ value before the collision; $\overline{B}$ is the typical average magnetic field strength in the reconnection region; $\left|V_z\right|$ is the magnitude of the initial velocity of each blob; $\alpha$ is the toroidal-to-poloidal flux ratio factor introduced in Equation (\ref{equ:B_phi}); $t_0$ and $L_0$ are the time and length normalization factors; $x_s$ is the initial misalignment between the center of the two blobs in the $x$ direction; $r_0$ is the initial radius of each blob before collision; $P$ is the uniform initial thermal pressure in the entire simulation box, and $\rho_{\rm bkg}$ is the uniform initial mass density outside the blobs. The density inside the blob is determined by $\sigma$.}
\end{table}

\begin{figure}[!htb]
\begin{center}
\includegraphics[angle=0,scale=0.55]{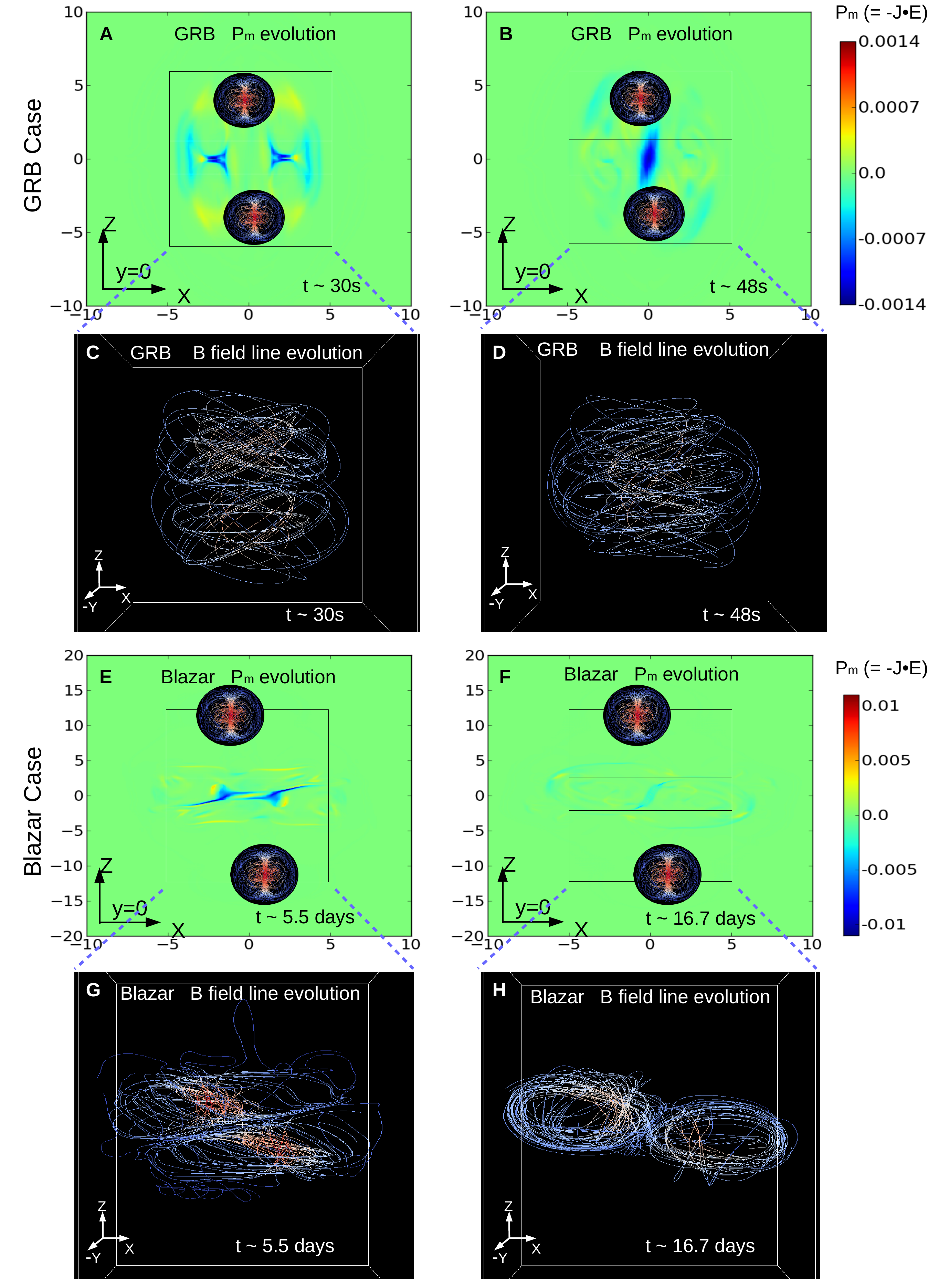}
\caption{The evolution of the quantity of $P_m=-\bld{J} \cdot \bld{E}$ in the $XZ$-plane (y=0) and the evolution of corresponding magnetic field line configuration. Panels ({\bf A}) and ({\bf B}) correspond to the $P_m$ evolution of GRB case ($P_m= -1$ corresponds to $2.4\times 10^4~{\rm  ergs \cdot s^{-1}\cdot cm^{-3}}$), and panels ({\bf E}) and ({\bf F}) are the $P_m$ evolution of blazar case ($P_m=-1$ corresponds to $1.05\times 10^{-6}~{\rm ergs \cdot s^{-1} \cdot cm^{-3}}$). The time marked in the panels roughly correspond to the time in Fig. \ref{fig:pol_GRB} \& \ref{fig:pol_Blazar}. The two magnetic blobs in each panel show the initial field line configuration and the rough initial positions of them. The blob size is enlarged for a better view. We define a reconnection region (inner black rectangle box) and a background region (outer black rectangle boxes) based on the $P_m$ distribution. Panels ({\bf C}, {\bf D}) and Panels ({\bf G}, {\bf H}) are the zoom-in magnetic field line configuration corresponding to each $P_m$ evolution panel of GRB and Blazar cases, respectively. The size of them roughly equals to the size of the background regions in $P_m$ evolution panels.}
\label{fig:power_combo}
\end{center}      
\end{figure}

We use $P_m=-\bld{J} \cdot \bld{E}$ as a normalization reference to inject relativistic particles\footnote{Due to the Poynting flux advection between simulation cells, $P_m$ is not the exact magnetic energy dissipation rate in each cell, but it is a good proxy for the magnetic energy dissipation to normalize the particle injection.}, where $\bld{J}$ is the current density, and $\bld{E}$ is the electric field strength. The negative value of $P_m$ implies magnetic energy dissipation. Figure \ref{fig:power_combo} shows two cuts of the $P_m$ evolution and the corresponding zoom-in magnetic field line configuration for the GRB and blazar cases, respectively.
The collision-induced magnetic reconnection happens in the contact area of the two blobs around the middle plane ($z=0$), where the magnetic energy dissipation mainly concentrates. Based on the distribution of the negative $P_m$ value, we define a reconnection region (inner black rectangle box) which is the major magnetic energy release region to produce strong flaring emission for GRB prompt emission and blazar flares. We also define a background region (outer black rectangle boxes) which contributes to the background emission. The magnetic field topology in the strong dissipation region evolves significantly between the two snapshots for both cases.

In each time step and each simulation cell with negative $P_m$ value, we inject non-thermal electrons with certain power-law distribution as follows:
\begin{equation}
n(\gamma_e)  =  \left \{
 \begin{array}{ll}
 n_0 (\gamma_e/\gamma_m)^{-1},& 10^2 < \gamma_e < \gamma_m,
\\
  n_0 (\gamma_e/\gamma_m)^{-3.2},& \gamma_m < \gamma_e < 10^6,
  \end{array}
   \right.
\label{eq:ele_dist_grb}
\end{equation}   
for the GRB jet \citep{uhm14,golenetskii10}, and
\begin{equation}
   n(\gamma_e)  =  n_0 \gamma_e^{-2}, ~~~~~~ 10^3 < \gamma_e < 5\times 10^4, 
\label{eq:ele_dist_blazar}
\end{equation}
for the blazar jet \citep{bottcher13}.
In the reconnection region, the normalization factor $n_0$ is calculated with the assumption that the injected particle energy rate is about half of the $P_m$ values. In the background region, on the other hand, $n_0$ is set to be uniform and normalized to the background emission level from the observations. The minimum Lorentz factor $\gamma_m$ is set to $1.4\times 10^4$ for the GRB case\footnote{This value is consistent with the analytical constraint of \cite{kumarcrumley15}.} to roughly match the observed peak energy ($E_p$) of GRB 100826A \citep{golenetskii10}.

\section{Radiation calculation}

Next, we use a 3D multi-zone polarization-dependent ray-tracing radiation code \citep{zhang14} named ``3DPol" to calculate the detailed radiation and polarization evolution properties based on the MHD simulations. The ``3DPol" code takes magnetic field and particle population as inputs, and calculates Stokes parameters at each spatial point. By adding up the polarized signals that arrive at the observer at the same time, the time-dependent PD and PA can be calculated. This code performs 
full 3D calculations of the polarization properties by taking into account detailed radiation and polarization transfer, light travel time effects, and time-, space- and frequency dependencies.

\section{Results}

\begin{figure}[!htb]
\begin{center}
\includegraphics[angle=0,scale=0.55]{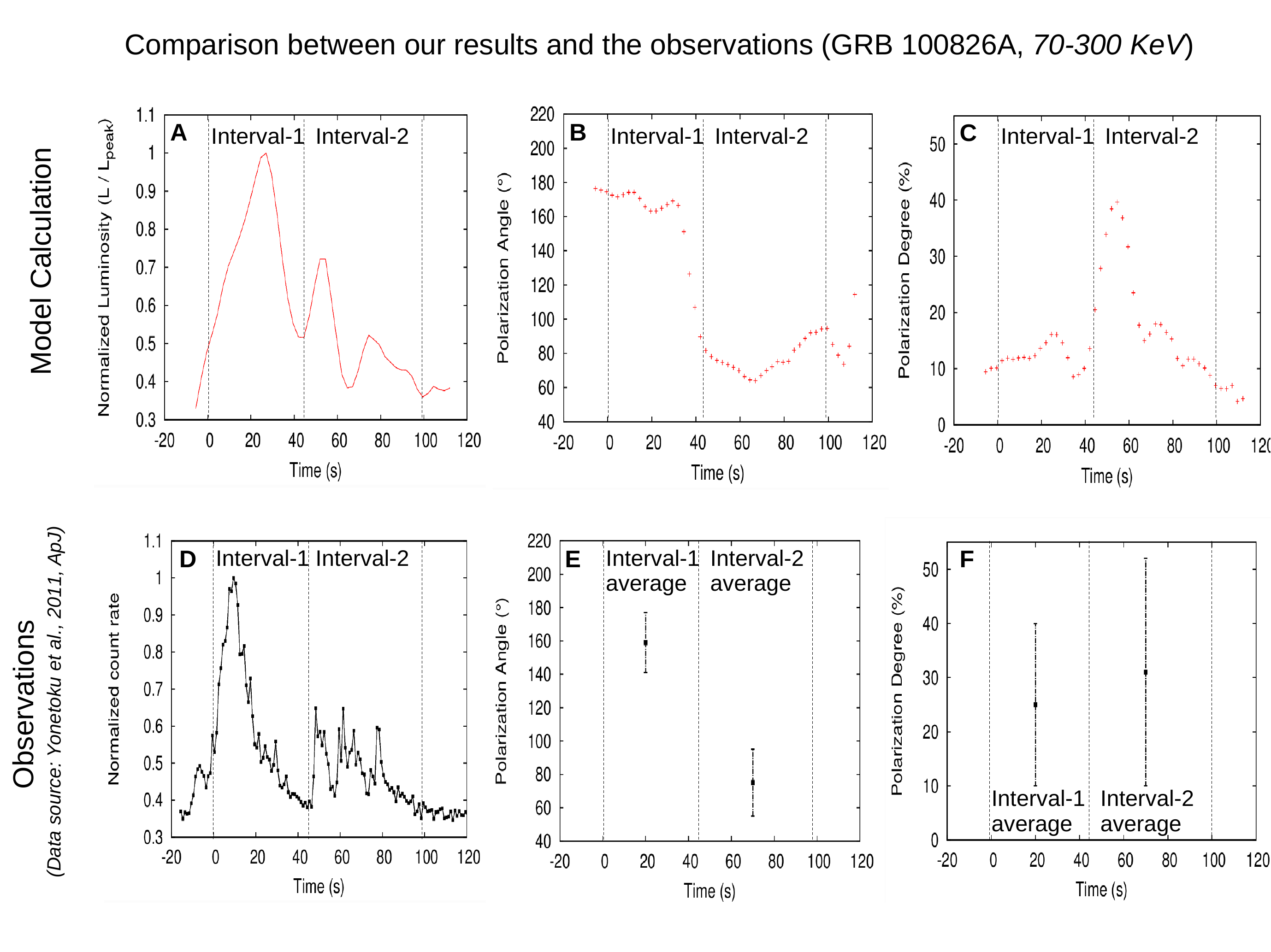}
\caption{The comparison between our simulation results and the observations of GRB 100826A in 70-300 KeV energy range. Panels ({\bf A}), ({\bf B}) and ({\bf C}) show our calculation results of the light curve, polarization angle (PA), and polarization degree (PD), respectively, whereas panels ({\bf D}), ({\bf E}) and ({\bf F}) are the corresponding observations \citep{yonetoku11} . The light curves of both observation and calculation are normalized by their peak values.
Following observational features are successfully reproduced from our model: 1. multiple global pulses; 2. the $90^{\circ}$ change of PA between the two intervals of the light curve; and 3. the trend of increasing PD between the two intervals.}
\label{fig:pol_GRB}
\end{center}      
\end{figure}

\begin{figure}[!htb]
\begin{center}
\includegraphics[angle=0,scale=0.55]{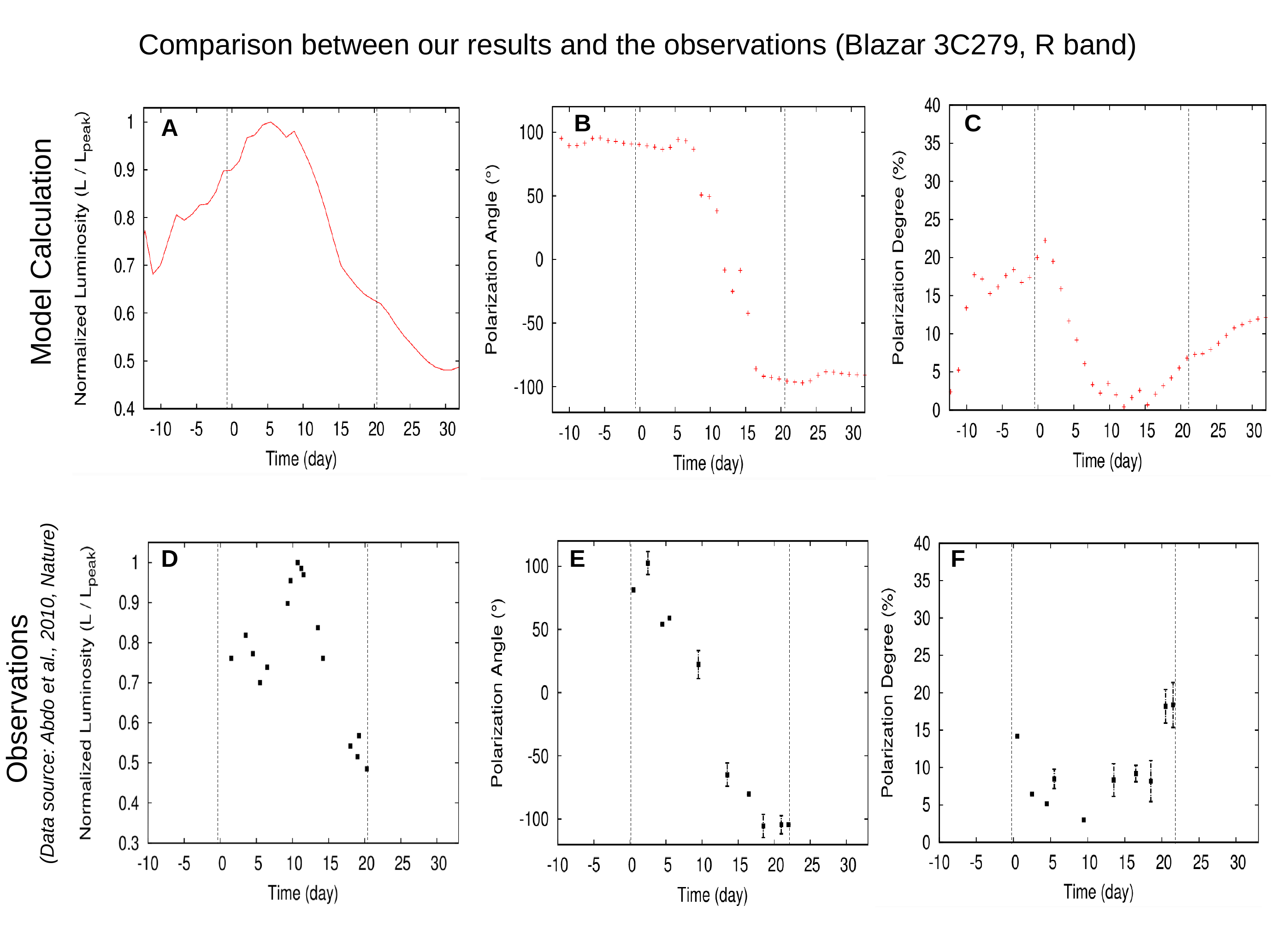}
\caption{The comparison between our simulation results and the observations of blazar 3C279 in optical R band. Panels ({\bf A}), ({\bf B}) and ({\bf C}) show our calculation results of the light curve, PA, and PD, respectively, whereas panels ({\bf D}), ({\bf E}) and ({\bf F}) are the corresponding observations \citep{abdo10}. The light curves of both observation and calculation are normalized by their peak values.
Following observational features are successfully reproduced from our model: 1. Single broad pulse light curve with a lower background after the flare;
2. $180^{\circ}$ PA angle swing; and 3. nearly zero PD around the flare's peak time ($t\sim 10$ days).}
\label{fig:pol_Blazar}
\end{center}      
\end{figure}

Our results and the referenced observations are shown in Fig. \ref{fig:pol_GRB} and Fig. \ref{fig:pol_Blazar} for the GRB and blazar cases, respectively. For the GRB case, the simulated light curve shows multiple global pulses (corresponding to the ``slow'' component of GRB emission \citep{gao12}), which is similar to the observations. The simulated PA on average rotates by $90^{\circ}$ between the two intervals of the light curves, which matches the observation quite well. In addition, the calculated PD also shows an increasing trend between the two intervals, consistent with the observations. For the blazar case, our results also match the observations well. In particular, the following three key features are reproduced: 1. The PA shows a $180^{\circ}$ angle swing; 2. The PD nearly drops to zero around the flare's peak time (around $t\sim 10$ days); 3. In the light curve, the background level after the flare is much lower than it was before the flare.

\section{Analyzes}

Our results can be understood in terms of the magnetic field configuration evolution during the collision processes. 
Since we assume that the observer views the system from the $x$-axis in the jet co-moving frame (the $1/\Gamma$ cone direction in the observer frame), only the radiation power from the $B_z$ and $B_y$ components contribute to the observed emission. The PA is roughly controlled by the dominant component of these two components, and the PD is related to the relative ratio between them. Since we inject the same energy distribution of non-thermal electrons in each cell, the synchrotron radiation powers of the two components in a certain region are related to $\sum_{i=1}^{n} n_0(i)*B_y(i)^2$ and $\sum_{i=1}^{n} n_0(i)*B_z(i)^2$, respectively, where $n$ is the total number of cells in this region, with $n_0$ introduced in Equations \ref{eq:ele_dist_grb} and \ref{eq:ele_dist_blazar}.

\begin{figure}[!htb]
\begin{center}
\includegraphics[angle=0,scale=0.55]{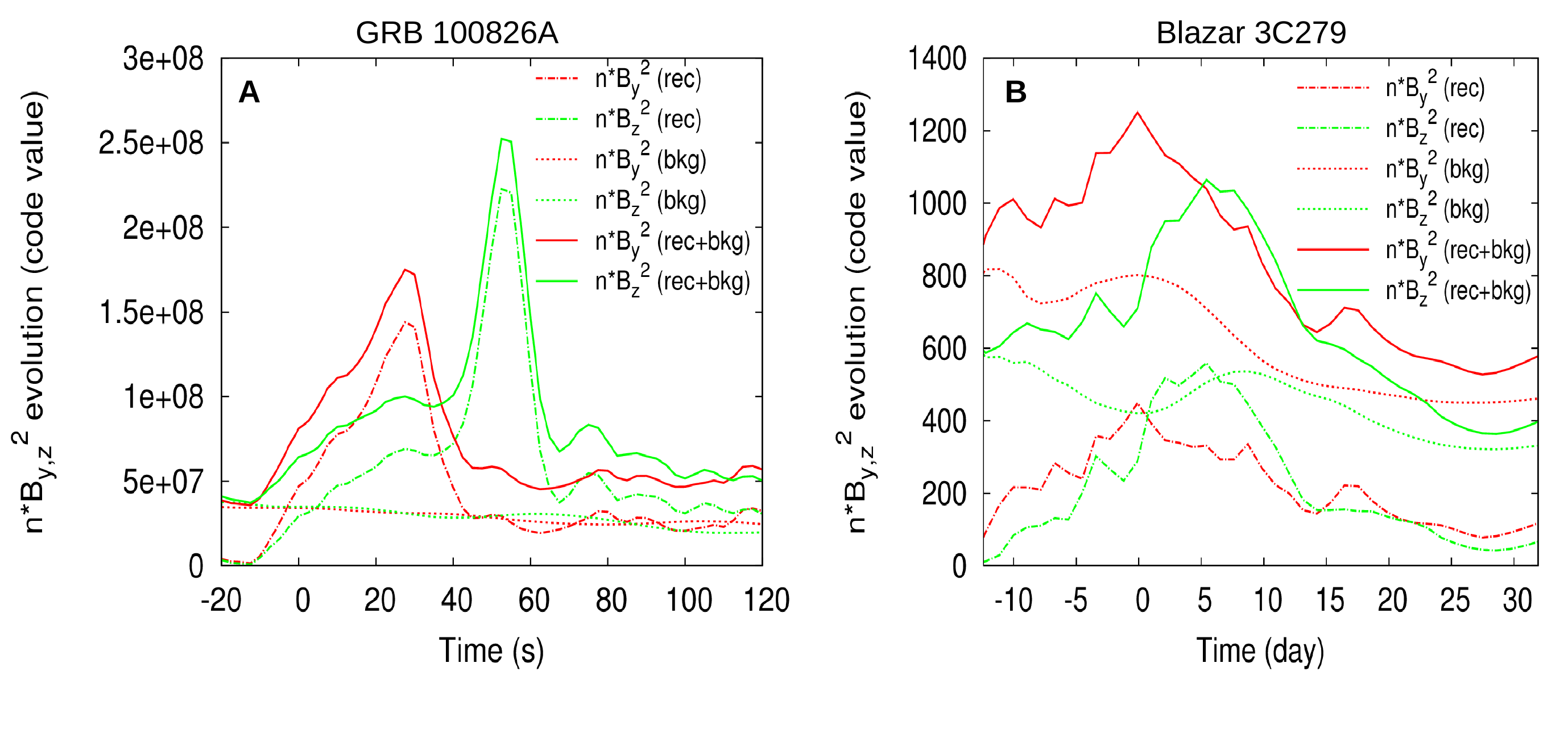}
\caption{The statistical evolution of the quantities $\sum_{i=1}^{n} n_0(i)*B_y(i)^2$ and $\sum_{i=1}^{n} n_0(i)*B_z(i)^2$ for the reconnection region (dash-dotted curves), the background region (dotted curves), and the superposition of them (solid curves), respectively. Panels ({\bf A}) and ({\bf B}) correspond to the GRB and blazar cases, respectively. }
\label{fig:power_evo_combo}
\end{center}      
\end{figure}

Figure \ref{fig:power_evo_combo} shows the evolution of the above two quantities for both cases. The final results should be controlled by the superposition (solid lines) from both the reconnection (dash-dotted lines) and background (dotted lines) regions. For the GRB case (left panel), the observation \citep{abdo10} shows a low level of the background emission, so we inject a small amount of non-thermal particles in the background region. Therefore, the radiation is dominated by the reconnection region. In this panel, there are two major power release intervals. The first one forms during the initial collision-induced strong magnetic reconnection process. Due to the small misalignment ($x_s$) and high-$\sigma$ environment, the blobs bounce back significantly. This process temporally suppresses energy release due to reconnection. After that, the imbalance due to the small misalignment is gradually built up, which triggers the later strong reconnection again to produce the second major flare (see detailed analysis of this process in Section 4.2 and Figure 12 in \cite{deng15}). This episodic power release process gives the multiple global pulses. On the other hand, the dominant component in the reconnection region switches from the $B_y$-dominated radiation to the $B_z$-dominated radiation between the two intervals, which is mainly due to the magnetic reconnection that reconnects the $B_y$ component to form the $B_z$ component gradually. This is the main reason to cause the $90^{\circ}$ PA change as observed.

For the blazar case (right panel), the observation \citep{abdo10} shows a relatively high level of background emission, so we inject more non-thermal particles in the background region to mimic the observation. Therefore, the superposition of the statistical quantities from the two regions is different from the GRB case. From the light curve point of view, due to the larger misalignment and moderate-$\sigma$ adopted in the simulation, the bouncing effect is relatively weak and more significant rotation is triggered, so that power release is relatively continuous in the reconnection region which gives rise to a single major flare in the light curve. The systematical decay of the magnetic field strength during the reconnection dissipation process causes the continuous decay of the background power release, which can explain the observed significantly lower level of the background emission after the flare than before. 
From the PA point of view, during the flare, the reconnection changes the dominant radiation component from $B_y$ to $B_z$ gradually. However at the late stage of the flare, $B_z$ domination becomes weaker and is even surpassed by the $B_y$ component again. This is mainly due to the relatively faster rotation caused by the larger misalignment, which rotates the system more than $45^{\circ}$ at the late stage. So $B_z$ becomes less dominating in the rotation co-moving frame, or even turns into $B_y$ domination for the fixed observer in the simulation frame. This double switching of the dominant component in the reconnection region causes the $180^{\circ}$ PA swing. 
From the PD point of view, the background contribution which is mainly $B_y$ dominant for this case is relatively high, so that the $B_z$ dominated emission from the reconnection region around the peak stage of the flare is nearly balanced by the $B_y$ dominated background emission. Therefore, the PD nearly drops to zero around the flare's peak time (around $t\sim 10$ days). We note that besides this representative case, many other blazars show significant PA swing along with relatively strong flare(s) \citep{blinov15}, which can be also potentially interpreted with our model. 

In realty, different GRB and Blazar events may have different parameters (e.g. $\sigma$, $\alpha$, $x_s$ etc.), or even different global magnetic field structures.
Based on our study, we find that as long as the change of magnetic field topology is mainly due to the magnetic reconnection process within a single global reconnection location,  the dominant magnetic field component would switch and give a significant (e.g. $90^{\circ}$) PA change. For more complicated cases involving multiple blobs colliding  and triggering many global reconnection events consecutively, the polarization evolution would be more complicated.

\section{Conclusion}

In summary, using MHD simulations, we are able to evolve the magnetic field topology of the GRB and blazer systems self-consistently by fully taking into account the collision-induced reconnection and rotation effect due to mis-alignment. The unified interpretation of the polarization properties of both GRBs and blazars suggest that similar underlying physics is in operation in black hole relativistic jet systems. In particular, collision-induced magnetic reconnection may hold the key to understand a variety of phenomenology.

\medskip
\acknowledgments
We thank the referee Pawan Kumar for helpful suggestions.
This work is supported by NASA through grants NNX15AK85G and NNX14AF85G, and by the LANL/LDRD program and Institutional Computing Programs at LANL and by DOE/Office of Fusion Energy Science through CMSO. 
We are grateful to Dr. Yonetoku for sharing the data with us.
We thank important technical support from Shengtai Li. 
We also thank helpful discussion and suggestions from Z. Lucas Uhm, Bin-Bin Zhang, Fan Guo, and Xiaocan Li.




\bigskip
\bigskip
\bigskip
\bigskip
\bigskip
\bigskip

\end{CJK*}
\end{document}